\documentstyle[12pt,fleqn,epsf]{article}

\def\Slash#1{{\rm\ooalign{\hfil$/$\hfil\crcr \hbox{$#1$}}}}
\newcommand\Kmbf[1]{\mbox{\boldmath{$#1$}}}

\begin{document}

\begin{flushright}
 TIT/HEP-405/NP
\end{flushright}

\begin{center}
{\Large \bf
Finite Quark Mass Effects in the Improved Ladder \par
Bethe--Salpeter Amplitudes} \par
\vskip 10mm
K. Naito\footnote{E-mail address: kenichi@th.phys.titech.ac.jp},
K. Yoshida, 
Y. Nemoto, 
M. Oka\\ 
{\it Department of Physics, Tokyo Institute of Technology,} \\
{\it Meguro, Tokyo 152-8551, Japan} 
\vskip 5mm
M. Takizawa\footnote{E-mail address: takizawa@ac.shoyaku.ac.jp}\\
{\it Laboratory of Computer Sciences, Showa College of Pharmaceutical Sciences,} \\
{\it Machida, Tokyo 194-8543, Japan}
\end{center}
\vskip 5mm
\begin{abstract}
\baselineskip 1.5pc
We study the finite quark mass effects of the low-energy QCD using 
the improved ladder Schwinger--Dyson and Bethe--Salpeter equations
which are derived in the manner consistent with 
the vector and axial-vector Ward--Takahashi identities. 
The non-perturbative mass-independent renormalization allows us to calculate
the quark condensate for a non-zero quark mass. 
We explicitly show that the PCAC relation holds.
The key ingredients are the Cornwall--Jackiw--Tomboulis effective action,
the generalized Noether current and the introduction of the regularization 
function to the Lagrangian. 
The reasonable values of the pion mass, the pion decay constant and the 
quark condensate are obtained with a rather large $\Lambda_{\rm QCD}$.
The pion mass square and the pion decay constant are almost proportional
to the current quark mass up to the strange quark mass region. 
It suggests that the chiral perturbation is applicable up to the strange
quark mass region.
 We study the validity of the approximation often used in solving the 
Bethe--Salpeter equations too.
\end{abstract}
\newpage 
\renewcommand{\thefootnote}{ \mbox{*} }
\section{Introduction} \label{SEC:H100319:1}
\hspace*{\parindent}
    The program to derive the observed properties of hadrons non-perturbatively
in QCD has been pursued with great intensity but not accomplished yet.
The concept of chiral symmetry and its spontaneous breakdown are among 
the most important aspects of low-energy hadron physics.  The spontaneous 
breakdown of chiral symmetry is believed to be responsible for a large 
part of the low-lying hadron masses as well as for the emergence 
of octet pseudoscalar mesons as Goldstone bosons.
In order to explain the observed hadron spectrum, one also needs 
small, explicitly chiral-symmetry breaking terms, namely, 
the flavor-dependent current quark mass terms.
\par
    The Conwall--Jackiw--Tomboulis (CJT) effective action approach 
\cite{CJT74} for composite operators is widely used to study 
the dynamical symmetry breaking phenomena in the quantum field 
theories.
The extremum condition for the effective action with respect to the quark 
propagator leads to the Schwinger--Dyson (SD) equation for the quark
propagator on the non-perturbative vacuum and the second variational 
derivative of the effective action with respect to the quark 
propagator leads to the Bethe--Salpeter (BS)
equation describing the bound states.
The advantage of the present approach is that the derived SD and BS equations
are consistent with the symmetry of the effective action evaluated in a 
certain approximation scheme.
\par
    The QCD SD equation for the quark propagator has been studied in the 
improved ladder approximation (ILA) by Higashijima \cite{H84} and Miransky
\cite{M84}. They took the ladder diagrams of one-gluon exchange 
between $q$ and $\bar{q}$ and assumed that the coupling constant is modified
according to the standard perturbative corrections. 
It has been shown that the asymptotic behavior of the solution is 
consistent with the leading order renormalization group analysis
while the infrared gluon exchange breaks chiral symmetry
dynamically. Aoki et al. solved the BS equation 
the $J^{PC} = 0^{-+}\,q\bar{q}$ state and confirmed the existence of the 
Nambu--Goldstone pion in this approximation \cite{KG1}.  
The numerical predictions of the pion decay constant $f_{\pi}$ and the 
quark condensate $\langle \overline{\psi} \psi \rangle$ are rather good. 
It was also shown that the BS amplitude shows the correct 
asymptotic behavior as predicted by the OPE in 
QCD \cite{ABKM91}.  The masses and decay constants for the lowest lying 
scalar, vector and axial-vector mesons have been evaluated by calculating 
the two point correlation functions for the composite operators 
$\overline{\psi} M \psi$.  The obtained values are in good agreement with the 
observed ones \cite{AKM91}. 
\par
 So far, the current quark mass term has not been introduced in the 
studies of the BS amplitudes in the ILA.  The purpose of the present 
paper is to investigate the effects of the finite current quark 
masses on the BS amplitudes for the $J^{PC} = 0^{-+}$ states.  
As shown in \cite{H84}, the asymptotic behavior of the solution of 
the SD equation for the quark propagator with the finite current 
quark mass is rather different from that in the chiral 
limit.  Therefore, it is important to study the effects of the finite
current quark masses not only on the SD equations but also on the BS
amplitudes.
\par
    There have been many studies of the pion BS amplitude using the 
effective models of QCD and/or the approximation schemes of 
the QCD \cite{RW94,MR97a}.
The advantages of the ILA model are as follows. 
(i) The model is given in the Lagrangian form so that one is able to
apply the CJT effective action formulation and study symmetry properties 
of the system.
(ii) The asymptotic behavior of the solutions of the SD and BS equations 
is consistent with the renormalization group analysis of QCD.
(iii) It has been shown that the ILA model corresponds to 
the local potential approximation with the ladder part in the non-perturbative
renormalization group approach \cite{AOKI97}. 
(iv) The angular integration in the SD equation can be performed analytically.
On the other hand, the disadvantages of the ILA model are as follows.
(i) The axial-vector Ward--Takahashi identity is violated \cite{KM92a}. 
(ii) The quark may not be confined in the color singlet state.
\par
    In the finite quark mass case, it has been known that there is a difficulty
in defining the quark condensate in the studies of the QCD SD equation. 
The extraction of the perturbative quark mass contribution in the UV region
is not sufficient to remove the UV divergence in the SD equation since the 
SD equation includes the non-perturbative contribution even in the UV region.
Inspired by Kusaka's idea of the non-perturbative renormalization of the 
fermion mass term in the mass-independent renormalization scheme \cite{kusaka},
we propose a novel way to renormalize the SD equation and define the quark 
condensate with the finite quark mass.
\par
    The paper is organized as follows. In Sec.~\ref{SEC:H100319:2} we explain 
the ILA model Lagrangian we have used in the present study. 
In Sec.~\ref{SEC:H100319:3} the SD equation is derived from the CJT action and 
the renormalization of the SD equation is discussed in 
Sec.~\ref{SEC:H100717:1}. In Sec.~\ref{SEC:H100319:4} our method of solving
the BS equation for the pseudoscalar meson is presented.
In Sec.~\ref{SEC:H100319:5} the formulation for the meson decay constant 
is given and the low-energy theorem is discussed.
Sec.~\ref {SEC:H100319:6} is devoted to the numerical results. 
Finally, summary and concluding remarks are given in Sec.~\ref{SEC:H100319:7}.
%
%
%
%
\section{Improved Ladder Model of QCD} \label{SEC:H100319:2}
    We work with the following Lagrangian density of the 
improved ladder approximation(ILA) model of QCD proposed by Aoki et al. 
\cite{KG1,AKM91},
\begin{eqnarray}
 {\cal L}[\psi,\overline{\psi}] & := & {\cal L}_{\rm free}[\psi,\overline{\psi}] + {\cal L}_{\rm int}[\psi,\overline{\psi}] \label{AEQ:H100319:1} \, ,\\
 {\cal L}_{\rm free}[\psi,\overline{\psi}] & := & \overline{\psi} f(\partial^2) (i\Slash{\partial}-m_0) \psi \, . \label{AEQ:H100319:2}
\end{eqnarray}
Here the function $f(\zeta)$ of $\zeta=\partial^2$ is introduced to provide 
a cut-off regularization of the ultraviolet divergences of the
quark loops.  
The reason we introduce the cut-off function at the Lagrangian level 
is to preserve the consistency between the SD and BS equations.
If one uses the regularization that is inconsistent between  
the SD and BS equations, the low-energy relation based on the chiral 
symmetry should be violated by the regularization. 
The function $f(\zeta)$ should satisfy $f(\zeta=0)=1$ and
$f(\zeta)\to \infty$ for $\zeta \gg \Lambda_{\rm UV}^2$.
In this paper, we employ the sharp cut-off function
\begin{equation}
 f(\zeta) = 1 + M \theta(\zeta-\Lambda_{\rm UV}^2),\quad M\to \infty.
 \label{AEQ:H100319:11}
\end{equation}
We introduce the bare mass of quarks $m_0$ which
is evaluated at $\Lambda_{\rm UV}$.\cite{KG1,NUCL1994}\,
In general $m_0$ is a diagonal flavor matrix i.e. 
$m_0={\rm diag}(m_u,m_d,m_s)$
for $N_f=3$. In this paper we deal only with a flavor independent mass 
and therefore the case with $SU(3)_F$ symmetry.
\par
    The interaction term is given by
\begin{eqnarray}
   {\cal L}_{\rm int}[\psi,\overline{\psi}](x) &:=& 
   -\frac{1}{2} \int_{pp'qq'} {\cal K}^{mm',nn'}(p,p';q,q') \nonumber \\
   & & {} \times \overline{\psi}_m(p)\psi_{m'}(p') 
   \overline{\psi}_n(q) \psi_{n'}(q') e^{-i(p+p'+q+q')x} \, ,
      \label{AEQ:H100319:3}
\end{eqnarray}
\begin{eqnarray}
   {\cal K}^{mm',nn'}(p,p';q,q') &=& \bar{g}^2 
   \left( (\frac{p_E-q_E'}{2})^2,(\frac{q_E-p_E'}{2})^2\right )  \nonumber \\
   &&\times iD^{\mu\nu} \left(\frac{p+p'}{2}-\frac{q+q'}{2} \right)
   (\gamma_\mu T^a)^{mm'} (\gamma_\nu T^a)^{nn'} 
      \label{AEQ:H100319:4}
\end{eqnarray}
where $\int_p$ denotes $\int \frac{d^4 p}{(2\pi)^4}$ and $p_E$ represents the
Euclidean momentum.
The Fourier transformations of fields are defined by
$\overline{\psi}(p)  =  \int d^4x e^{ipx}\overline{\psi}(x)$ and
$\psi(p)  =  \int d^4x e^{ipx}\psi(x)$.
The indices $m,n,\cdots$
are combined indices $m:=(a,i,f),\,n:=(b,j,g),\cdots$ with Dirac indices
$a,b,\cdots$ and color indices $i,j,\cdots $ and flavor indices $f,g,\cdots$.
$T^a$ denotes the generator of the color $SU(N_C)$. 
According to Higashijima and Miransky, we choose a particular set of the
momenta that determines the running coupling constant, i.e.,
\begin{equation}
   \bar{g}^2(p_E^2,q_E^2) = \theta(p_E^2-q_E^2) g^2(p_E^2) + 
   \theta(q_E^2-p_E^2) g^2(q_E^2). 
   \label{AEQ:H100319:7}
\end{equation}
This way of introducing the running coupling constant is very natural from the 
non-perturbative renormalization group approach with the local potential
approximation \cite{AOKI97}.
It is often called the Higashijima-Miransky approximation.
The infrared cut-off $t_{\rm IF}$ is introduced in the running coupling 
constant as
\begin{equation}
 g^2(p_E^2) := \left\{ 
   \begin{array}{ll} \displaystyle{\frac{1}{\beta_0}\frac{1}{1+t}} &
   \mbox{ for } t_{\rm IF}\le t \\ 
   \mbox{} & \\
   \displaystyle{\frac{1}{2\beta_0}\frac{1}{(1+t_{\rm IF})^2}
   \left[3t_{\rm IF}-t_0+2-\frac{(t-t_0)^2}{t_{\rm IF}-t_0}\right]} &
   \mbox{ for } t_0\le t \le t_{\rm IF} \\
   \mbox{} & \\
   \displaystyle{\frac{1}{2\beta_0}\frac{3t_{\rm IF}-t_0+2}{(1+t_{\rm IF})^2}}&
   \mbox{ for } t\le t_0 \end{array} \right. ,
   \label{AEQ:H100319:8}
\end{equation}
\begin{equation}
 t := \ln \frac{p_E^2}{\Lambda_{\rm QCD}^2} - 1 ,
 \label{AEQ:H100319:9}
\end{equation}
\begin{equation}
 \beta_0 := \frac{1}{(4\pi)^2}\frac{11N_C-2N_f}{3}. \label{AEQ:H100319:10}
\end{equation}
Above $t_{\rm IF}$, $g^2(p_E^2)$ develops according to the one-loop 
result of the 
QCD renormalization group equation and below $t_0$, $g^2(p_E^2)$ is kept 
constant. These two regions are connected by the quadratic polynomial so that
$g^2(p_E^2)$ becomes a smooth function.  Here $N_C$ is the number of colors 
and $N_f$ is the number of active flavors. We use $N_C = N_f = 3$ in our 
numerical studies.
The gluon propagator is given in the Landau gauge
\begin{equation}
 iD^{\mu\nu}(k)  = \left( g^{\mu\nu} - \frac{k^\mu k^\nu}{k^2} \right)
 \frac{-1}{k^2}.
 \label{AEQ:H100319:10a}
\end{equation}
%
%
\section{SD equation} \label{SEC:H100319:3}
   In order to derive the Schwinger--Dyson (SD) equation, 
we use the formalism of the Cornwall--Jackiw--Tomboulis (CJT) 
effective action \cite{CJT74} which is given by 
\begin{equation}
 \Gamma[S_F] := i {\rm Tr}{\rm Ln}[S_F] - i{\rm Tr}[S_0^{-1}S_F] 
 + \Gamma_{\rm loop}[S_F].
 \label{AEQ:H100319:12}
\end{equation}
 The last term of Eq.$(\ref{AEQ:H100319:12})$ is the residual term.
 Multiplying a factor $i$, $i\Gamma_{\rm loop}[S_F]$ is given by the sum of
 all Feynman amplitudes of 2-loop or higher-loop 2-particle irreducible
 vacuum diagrams in which every bare quark propagator
\begin{equation}
 S_0(x,y) = \int_q e^{-iq(x-y)} \frac{1}{f(-q^2)} \frac{i}{\Slash{q}-m_0}
 \label{AEQ:H100319:13}
\end{equation}
is replaced by the full one 
\begin{equation}
 S_F(x,y) = \langle 0 | T \psi(x) \overline{\psi}(y) | 0 \rangle. 
 \label{AEQ:H100319:14}
\end{equation}
The SD equation is the stability condition of the CJT action
\begin{equation}
 \frac{\delta \Gamma[S_F]}{\delta S_{Fmn}(x,y)} = 0.
  \label{AEQ:H100319:15}
\end{equation}
Throughout this paper, we employ the lowest order (lowest--loop)
expansion of the $\Gamma_{\rm loop}[S_F]$ as
\begin{eqnarray}
 \Gamma_{\rm loop}[S_F] & = & -\frac{1}{2}\int d^4 x {\cal K}^{m_1m_2,n_1n_2}
 \left( i\partial_{x_1},i\partial_{x_2};i\partial_{y_1},i\partial_{y_2}\right)
  \label{AEQ:H100322:1}\\
 & & {} \times \left[ S_{Fm_2m_1}(x_2,x_1)S_{Fn_2n_1}(y_2,y_1)
 -S_{Fm_2n_1}(x_2,y_1)S_{Fn_2m_1}(y_2,x_1)\right]\Big|_* \nonumber 
\end{eqnarray}
where the symbol $*$ means to taking $x_1,x_2,y_1,y_2\to x$ after all 
the derivatives are operated.
This leads to the ILA model, where the SD equation is given in momentum 
space by
\begin{equation}
 iS_F^{-1}(q) - iS_0^{-1}(q) + C_F \int_p \bar{g}^2(q_E^2,p_E^2)
  iD^{\mu\nu}(p-q)\gamma_\mu S_F(p) \gamma_\nu = 0 \, ,
 \label{AEQ:H100319:16}
\end{equation}
with
\begin{equation}
  C_F = \frac{{\rm tr} \left[T^a T^a \right]}{N_C} = \frac{N_C^2 - 1}{2 N_C} \, .
  \label{AEQ:H100718:1}
\end{equation}
Now we introduce the regularized propagator as
\begin{eqnarray}
 S_0^R(q) & := & f(-q^2) S_0(q) = \frac{i}{\Slash{q}-m_0},
  \label{AEQ:H100319:17}\\
 S_F^R(q) & := & f(-q^2) S_F(q). \label{AEQ:H100319:18}
\end{eqnarray}
Then the SD equation $(\ref{AEQ:H100319:16})$ becomes
\begin{equation}
 i{S_F^R}^{-1}(q) -i {S_0^R}^{-1}(q) + \frac{C_F}{f(-q^2)} 
 \int_p \frac{1}{f(-p^2)} \bar{g}^2(q_E^2,p_E^2) 
  iD^{\mu\nu}(p-q)\gamma_\mu S^R_F(p) \gamma_\nu = 0
 \label{AEQ:H100319:19}
\end{equation}
in which one finds that the integral is cut-off 
at $p_E^2 = -p^2 = \Lambda_{\rm UV}^2$ 
due to the function $f(-p^2)$.
Substituting the general form of the SD solution
\begin{equation}
 S_F^R(q) = \frac{i}{A(q^2)\Slash{q}-B(q^2)} \label{AEQ:H100319:20}
\end{equation}
we obtain a set of integral equations
\begin{eqnarray}
  A(q^2)  & = & 1 +  \frac{iC_F}{q^2f(-q^2)} \int_p 
  \frac{\bar{g}^2(q_E^2,p_E^2)}{f(-p^2)}
  \frac{3(p^2+q^2)(pq)-4(pq)^2-2p^2q^2}{(q-p)^4} \nonumber\\
&& \times \frac{A(p^2)}{p^2 A^2(p^2)-B^2(p^2)}, \label{AEQ:H100319:21} 
\end{eqnarray}
\begin{equation}
 B(q^2) = m_0+\frac{iC_F}{f(-q^2)} \int_p
 \frac{\bar{g}^2(q_E^2,p_E^2)}{f(-p^2)} \frac{1}{(p-q)^2}
  \frac{-3 B(p^2)}{p^2 A^2(p^2)-B^2(p^2)}. \label{AEQ:H100319:22} 
\end{equation}
After the Wick rotation, we obtain
\begin{equation}
 A(-q_E^2) \equiv 1
 \label{AEQ:H100319:23}
\end{equation}
from Eq.$(\ref{AEQ:H100319:21})$. 
This is another advantage of the Higashijima--Miransky approximation,
where the running coupling constant is defined so as to make 
the wave function renormalization $Z_2$ unity \cite{KG1,KM92a}.
Then we find an integral equation for $B(-q_E^2)$ as
\begin{equation}
 B(-q_E^2) = m_0 + \frac{3C_F}{16\pi^2} \int_0^{\Lambda^2_{\rm UV}} d p_E^2 \,
 \bar{g}^2(q_E^2,p_E^2) \frac{p_E^2}{\max\{q_E^2,p_E^2\}} 
 \frac{B(-p_E^2)}{ p_E^2 + B^2(-p_E^2)}. 
 \label{AEQ:H100319:24}
\end{equation}
%
%
%
%
%
\section{Renormalization of quark mass} \label{SEC:H100717:1}
    In this section we discuss the renormalization of the 
quark mass.  The operator product expansion 
analysis shows that in the asymptotic region the QCD quark mass function 
$B(-q_E)$ for three quark flavors behaves as follows \cite{POLITZER76}.
\begin{equation}
  B(-q_E^2) = m_R(\mu^2) \, \left[ \frac{g^2(q_E^2)}{g^2(\mu^2)} \right]^{4/9} 
  - \xi_R (\mu^2) \, 
  \frac{g^2(q_E^2)}{3 q_E^2} \,
  \left[ \frac{g^2(q_E^2)}{g^2(\mu^2)} \right]^{-4/9} \, ,
  \label{AEQ:H100721:1}
\end{equation}
where $m_R(\mu^2)$ is the current quark mass renormalized at $\mu^2$ 
and $\xi_R(\mu^2) := \langle \overline{\psi} \psi \rangle_R$ is the quark 
condensate renormalized at $\mu^2$. 
The improved ladder model of QCD is the model which is constructed so as to 
reproduce the QCD asymptotic behavior.  Therefore we introduce the
renormalization condition of the quark mass so that  
the solution of the SD equation can be interpreted as the QCD quark 
mass function in the asymptotic region.
\par
    The quark mass function calculated in the effective model of QCD 
can be expressed in the similar fashion as Eq.(\ref{AEQ:H100721:1}) 
\begin{equation}
  B(-q_E^2) = m_R(\mu^2) \, F(q_E^2, \mu^2) - 
  \xi_R (\mu^2) \, G(q_E^2, \mu^2) \,.
  \label{AEQ:H100721:2}
\end{equation}
Then we introduce the renormalization condition 
\begin{equation}
   F(\mu^2, \mu^2) = 1 \,,
   \label{AEQ:H100721:3}
\end{equation}
which is equivalent to 
\begin{equation}
   \left. \frac{\partial B(-\mu^2)}{\partial m_R(\mu^2)} 
   \right|_{m_R(\mu^2)=0} = 1 \, .
   \label{AEQ:H100721:4}
\end{equation}
This mass independent renormalization condition for the SD equation is 
first proposed by Kusaka \cite{kusaka}.
The mass renormalization constant $Z_m = Z_m(\Lambda_{\rm UV}^2, \mu^2)$ 
is introduced by 
\begin{equation}
   m_0(\Lambda_{\rm UV}^2) = Z_m^{-1} m_R(\mu^2) \, .
   \label{AEQ:H100721:5} 
\end{equation}
It should be noted here that the renormalization constant is independent
of mass in this renormalization scheme.  It will be explicitly 
shown later in this section.  By substituting $m_0$ by Eq.(\ref{AEQ:H100721:5}),
the SD equation (\ref{AEQ:H100319:24}) becomes
\begin{equation}
 B(-q_E^2) = Z_m^{-1}\, m_R(\mu^2) + \int_0^{\Lambda^2_{\rm UV}} d p_E^2 \,
 K(q_E^2,p_E^2) \, \frac{B(-p_E^2)}{ p_E^2 + B^2(-p_E^2)} \, , 
 \label{AEQ:H100721:6}
\end{equation}
with
\begin{equation}
   K(q_E^2,p_E^2) := \frac{3C_F}{16\pi^2} \,
   \bar{g}^2(q_E^2,p_E^2) \frac{p_E^2}{\max\{q_E^2,p_E^2\}} \, .
   \label{AEQ:H100721:7}
\end{equation}
By differentiating this equation with respect to $m_R(\mu^2)$ and 
taking $m_R(\mu^2) = 0$, one obtains 
\begin{eqnarray}
 \left. \frac{\partial B(-q_E^2)}{\partial m_R(\mu^2)} 
 \right|_{m_R(\mu^2)=0} & = & Z_m^{-1}\, \nonumber \\ 
 && + \int_0^{\Lambda^2_{\rm UV}} d p_E^2 \, K(q_E^2,p_E^2) \, 
 \frac{p_E^2 - B_0^2(-p_E^2)}{\left( p_E^2 + B_0^2(-p_E^2)\right)^2} \, 
 \left. \frac{\partial B(-p_E^2)}{\partial m_R(\mu^2)} 
 \right|_{m_R(\mu^2)=0}  . 
 \label{AEQ:H100722:1}
\end{eqnarray}
The renormalization condition (\ref{AEQ:H100721:4}) leads to
\begin{equation}
  Z_m^{-1} = 1 - \int_0^{\Lambda^2_{\rm UV}} d p_E^2 \, K(\mu^2,p_E^2) \, 
 \frac{p_E^2 - B_0^2(-p_E^2)}{\left( p_E^2 + B_0^2(-p_E^2)\right)^2} \, 
 \left. \frac{\partial B(-p_E^2)}{\partial m_R(\mu^2)} 
 \right|_{m_R(\mu^2)=0}  . 
 \label{AEQ:H100722:2}
\end{equation}
Here $B_0(-q_E^2)$ is the solution of the SD equation in the chiral limit,
namely,
\begin{equation}
 B_0(-q_E^2) =  \frac{3C_F}{16\pi^2} \int_0^{\Lambda^2_{\rm UV}} d p_E^2 \,
 \bar{g}^2(q_E^2,p_E^2) \, \frac{p_E^2}{\max\{q_E^2,p_E^2\}} \,
 \frac{B_0(-p_E^2)}{ p_E^2 + B_0^2(-p_E^2)} \, . 
 \label{AEQ:H100722:3}
\end{equation}
Now the combination of Eqs.(\ref{AEQ:H100722:1}) and (\ref{AEQ:H100722:2})
yields an integral equation for 
$\left. \frac{\partial B(-q_E^2)}{\partial m_R(\mu^2)} \right|_{m_R(\mu^2) = 0}$.
Eq.(\ref{AEQ:H100722:2}) explicitly shows that the mass renormalization 
constant $Z_m$ does not depend on the quark mass.
\par
    In order to obtain the quark mass function with the renormalized 
current quark mass, one first calculates $B_0(-q_E^2)$ by solving the SD 
equation in the chiral limit Eq.(\ref{AEQ:H100722:3}). 
Next $Z_m$ and
$\left.\frac{\partial B(-q_E^2)}{\partial m_R(\mu^2)} \right|_{m_R(\mu^2) = 0}$
are obtained by solving Eqs.(\ref{AEQ:H100722:1}) and Eq.(\ref{AEQ:H100722:2}).
Finally Eq.(\ref{AEQ:H100721:6}) is solved to find $B(-q_E^2)$.
\par
    Let us now propose the following definition of the quark condensate.
\begin{equation}
   \xi_R(\mu^2) := 
   Z_m^{-1} \xi_0(\Lambda_{\rm UV}^2) \, ,
   \label{AEQ:H100722:4}
\end{equation}
\begin{equation}
   \xi_0(\Lambda_{\rm UV}^2) :=
   -\, \left( \int^{\Lambda_{\rm UV}} \frac{d^4 q}{(2\pi)^4} {\rm tr}[S_F^R(q)]
   -\, \int^{\Lambda_{\rm UV}} \frac{d^4 q}{(2\pi)^4} 
   {\rm tr}[S_F^R(q)_{\rm pert}]
    \right) \, ,
    \label{AEQ:H100722:5}
\end{equation}
\begin{equation}
   S_F^R(q) := \frac{i}{\Slash{q} - B(q^2)} \, ,
   \label{AEQ:H100722:6}
\end{equation}
\begin{equation}
   S_F^R(q)_{\rm pert} := 
   \left. \frac{\partial S_F^R(q)}{\partial m_R(\mu^2)} \right|_{m_R(\mu^2)=0}
   \, m_R(\mu^2) \, .
   \label{AEQ:H100722:7}
\end{equation}
In order to avoid the divergence originated by the perturbative quark mass
contribution to the quark condensate in the UV region, the perturbative
quark mass contribution should be subtracted.  The key point of our 
definition of the quark condensate is Eq.(\ref{AEQ:H100722:7}), namely,
the perturbative quark mass contribution is defined using the fully 
calculated quark mass function $B(q^2)$.  The subtraction of the 
perturbative quark mass contribution obtained by the operator product
expansion approach in the UV region is not sufficient.
Our definition of the quark condensate has the desirable property:
\begin{equation}
  m_0(\Lambda_{\rm UV}^2)\,
  \xi_0(\Lambda_{\rm UV}^2) = 
  m_R(\mu^2) \,
  \xi_R(\mu^2) \, .
  \label{AEQ:H100722:8}
\end{equation}
%
%
\section{BS equation for Pseudoscalar Mesons} \label{SEC:H100319:4}
 The homogeneous BS equation is given by
\begin{equation}
 \frac{\delta^2 \Gamma[S_F]}{\delta S_{Fmn}(x,y) \delta S_{Fn'm'}(y',x')}
  \chi_{n'm'}(y',x';P_B) = 0 \, .
 \label{AEQ:H100319:25}
\end{equation}
Here the BS  amplitude is defined by
\begin{equation}
 \chi_{nm}(y,x;P_B) := \langle 0 | T \psi_n(y) \overline{\psi}_m(x) |
 \Kmbf{P}_B \rangle 
\label{AEQ:H100319:26}
\end{equation}
for a $q$-$\bar{q}$ state $|\Kmbf{P}_B\rangle$. 
The normalization condition is 
$\langle \Kmbf{P}_B |  \Kmbf{P}'_B \rangle = 
(2\pi)^3 2P_{B0}\delta^3(\Kmbf{P}_B - \Kmbf{P}'_B)$ and
$P_B:=(\sqrt{M_B^2+\Kmbf{P}_B^2},\Kmbf{P}_B)$ is the on-shell
momentum.  
Eq.$(\ref{AEQ:H100319:25})$ is expressed in momentum space
\begin{equation}
 S^{-1}_{F}(q_+) \chi(q;P_B) S^{-1}_{F}(q_-) = -i C_F
 \int_k \bar{g}^2(q_E^2,k_E^2) iD^{\mu\nu}(q-k) \gamma_\mu \chi(k;P_B) 
 \gamma_\nu \, , \label{AEQ:H100320:1}
\end{equation}
with
\begin{equation}
 q_+ = q+\frac{P_B}{2},\quad q_- = q-\frac{P_B}{2} \, ,
 \label{AEQ:H100320:1a}
\end{equation}
where the Fourier transformation of the BS amplitude is defined by
\begin{equation}
 \chi_{nm}(y,x;P_B) = e^{-iP_BX} \int_q e^{-iq(y-x)}\chi_{nm}(q;P_B)
 ,\, \quad X=\frac{y+x}{2}. \label{AEQ:H100320:2}
\end{equation}
We introduce the regularized BS amplitude by
\begin{equation}
 \chi^R_{nm}(q;P) := f(-q_+^2) \chi_{nm}(q;P) f(-q_-^2),
  \label{AEQ:H100320:3}
\end{equation}
then Eq.$(\ref{AEQ:H100320:1})$ is rewritten as
\begin{eqnarray}
\lefteqn{
 {S^R_F}^{-1}(q_+) \chi^R(q;P_B) {S^R_F}^{-1}(q_-) } \nonumber \\
 & = & -i C_F
 \int_k \frac{1}{f(-k_+^2)f(-k_-^2)} 
 \bar{g}^2(q_E^2,k_E^2) iD^{\mu\nu}(q-k) \gamma_\mu \chi^R(k;P_B)
 \gamma_\nu. \label{AEQ:H100320:4}
\end{eqnarray}
We see again that the integral equation is regularized correctly.
\par
    The BS amplitude for the pseudoscalar meson can be written
in terms of four scalar amplitudes,
\begin{eqnarray}
 \lefteqn{ 
 \chi^R_{nm}(k;P) = \delta_{ji} \frac{(\lambda^{\alpha})_{gf}}{2}\bigg[
  \bigg(\phi_S(k;P) + \phi_P(k;P) \Slash{k} + \phi_Q(k;P)\Slash{P} 
  } \nonumber \\
  & & \quad {}  +\frac{1}{2}\phi_T(k;P)(\Slash{P}\Slash{k}-
  \Slash{k}\Slash{P})\bigg)\gamma_5\bigg]_{ba}
 \label{AEQ:H100320:5}
\end{eqnarray}
$
$
where $\lambda^{\alpha}$ denotes the flavor matrix.
Substituting this into Eq.$(\ref{AEQ:H100320:4})$, we obtain coupled 
integral equations for four scalar amplitudes. The explicit form is rather
complicated and given in appendix. The integral equations
can be written down formally
\begin{equation}
 \phi_{\cal A}(q;P_B) = \int_k {\cal M}_{{\cal AB}}(q,k;P_B)
  \phi_{\cal B}(k;P_B)
 \label{AEQ:H100320:6}
\end{equation}
where ${\cal A},{\cal B}$ denotes $S,P,Q,T$.
Among the four dimensional integration $d^4 k$, two of the integrations 
can be performed analytically after the Wick rotation and 
we set the total momentum $P_{BE}=(M_E,0,0,0)$. 
Then we obtain
\begin{equation}
 \phi_{\cal A}(q_R,q_\theta;M_E) = \int_{(k_R,k_\theta)\in I} d k_R 
dk_\theta {\cal M}_{\cal AB}
 (q_R,q_\theta;k_R,k_\theta;M_E) \phi_{\cal B}(k_R,k_\theta;M_E) 
 \label{AEQ:H100320:7}
\end{equation}
where
\begin{equation}
 k_E^2=k_R^2,\quad k_EP_E = k_R M_E \sin k_\theta
 \label{AEQ:H100320:8}
\end{equation}
and the integral region is given  by
\begin{equation}
 I := \left\{ (k_R,k_\theta) \,\bigg|\,  k_R^2 \pm k_RM_E\sin k_\theta + \frac{M_E^2}{4}
 \le \Lambda_{\rm UV}^2 \right\}.
 \label{AEQ:H100320:9}
\end{equation}
This integral region is determined uniquely by the cut-off function in 
Eq.$(\ref{AEQ:H100320:4})$ and is consistent with the SD equation
$(\ref{AEQ:H100319:24})$.  
\par
    There is no solution of Eq.$(\ref{AEQ:H100320:7})$ for a real $M_E$ 
because $M_E$ is a Euclidean meson mass whose square is negative
$M_E^2=-M_B^2$. Since the SD equation can be solved only for 
space like region $q_E^2 \ge 0$, the region $M_E^2<0$ is not accessible.
Instead of solving Eq.$(\ref{AEQ:H100320:7})$,
we convert it into an eigenvalue equation for a fixed $M_E^2\ge 0$,
given by
\begin{equation}
 \lambda \phi_{\cal A}(q_R,q_\theta;M_E) = \int d k_R dk_\theta
  {\cal M}_{\cal AB}(q_R,q_\theta;k_R,k_\theta;M_E) 
  \phi_{\cal B}(k_R,k_\theta;M_E) 
 \label{AEQ:H100320:10}
\end{equation}
where $\lambda$ is the eigenvalue that is equal to unity for the
 solutions of Eq.$(\ref{AEQ:H100320:7})$. 
We solve Eq.$(\ref{AEQ:H100320:10})$ numerically using 
the iteration procedure.
When we iterate Eq.$(\ref{AEQ:H100320:10})$, the eigen-function 
associated with the maximum absolute eigenvalue is dominated.
Then we obtain the maximum eigenvalue and its eigen-function.
\par
  In the numerical calculations, we use the discretization
of the continuous variable $(q_R,q_\theta)$ and come across
a problem that the kernel $K_{\cal AB}$ diverges at the point
$(q_R,q_\theta)=(k_R,k_\theta)$.
This divergence, which is originated from the gluon propagator,
does not cause a real divergence.
We may remove this divergence by carefully choosing  
the discretization points in the iteration procedure.
\par
    Once we obtain (the largest absolute) $\lambda$ as a function of 
$M_E^2\ge0$, then we extrapolate $\lambda(M_E^2)$ to the time-like 
region $M_E^2<0$ and
look for the on-shell point where $\lambda(-M_B^2)=1$.
Since this extrapolation is the most ambiguous procedure in our calculation,
we will later consider another function which is similarly 
extrapolated to the one-shell value and compare the predicted values 
$M_B^2$ obtained in the two independent extrapolations.
\par

\section{Decay Constant and low-energy relation} \label{SEC:H100319:5}
 To obtain the decay constant, we need the normalization of the BS amplitude.
The normalization condition of the BS amplitude is derived from the
inhomogeneous BS equation
\begin{equation}
 \frac{1}{i} \frac{\delta ^2 \Gamma[S_F]}{\delta S_{Fmn}(x,y) \delta
  S_{Fn'm'}(y',x')} G_{C;n'm'm''n''}^{(2)}(y'x';x''y'') =
   \delta_{m''m}\delta_{nn''}\delta(x''-x)\delta(y-y'')
 \label{AEQ:H100407:2}
\end{equation}
where
\begin{eqnarray}
 G_{C;nmm'n'}^{(2)}(yx;x'y') & := & \langle 0 | T \psi_n(y)
 \overline{\psi}_m(x)
 \psi_{m'}(x') \overline{\psi}_{n'}(y') | 0 \rangle  \nonumber \\
& & {} \quad - \langle 0 | T\psi_n(y)\overline{\psi}_m(x) |
 0 \rangle \langle 0 | T\psi_{m'}(x')\overline{\psi}_{n'}(y')
  | 0 \rangle. \label{AEQ:H100407:3}
\end{eqnarray}
In the momentum space, Eq.$(\ref{AEQ:H100407:2})$ gives
\begin{eqnarray}
\lefteqn{ i \int_q \frac{1}{f(-q_+^2)f(-q_-^2)}\chi^R_{n_1m_1}(q;P_B)
\overline{\chi}^R_{m_2n_2}(q;P_B)\frac{\partial}{\partial P^\mu}\Big(
 S^{R-1}_{Fn_2n_1}(q_+) S^{R-1}_{Fm_1m_2}(q_-) \Big) }  \nonumber \\ 
& & {} + i\int_q \frac{-(q_+)_\mu f'(-q_+^2)f(-q_-^2)-(q_-)_\mu 
f'(-q_-^2)f(-q_+^2) }{f^2(-q_+^2) f^2(-q_-^2) } \nonumber \\
& & \quad {} \times \chi^R_{n_1m_1}(q;P_B)S^{R-1}_{Fm_1m_2}(q_-) 
\overline{\chi}^R_{m_2n_2}(q;P_B) S^{R-1}_{Fn_2n_1}(q_+)
=-2P_\mu,\quad P \to P_B. 
 \label{AEQ:H100407:4}
\end{eqnarray}
In the case of the sharp cut-off function $(\ref{AEQ:H100319:11})$,
the second term in the LHS of Eq.$(\ref{AEQ:H100407:4})$ 
does not contribute and the integral region in the first term 
is determined uniquely.
\par
     Let us now turn to the discussion of the axial-vector 
Ward--Takahashi (WT) identity.  It has been found that the 
axial-vector WT identity is violated in the Higashijima-Miransky
(HM) approximation \cite{JM91}.  Of course the Goldstone theorem 
holds in this case because the HM approximation respects the global
chiral symmetry.  The chiral WT identity in the ladder approximation 
has been carefully studied in \cite{KM92a}.  The reason of the violation 
of the axial-vector WT identity is that the HM approximation breaks local 
chiral symmetry.  As shown in \cite{KM92a}, the improved ladder 
approximation of the SD and BS equations preserves the WT identity for 
the axial-vector vertex if and only if one uses the gluon momentum 
square as the argument of the running coupling constant.  However,
in this case renormalization factor $Z_2$ of the quark wave function 
deviates from unity in the Landau gauge.  In order to avoid such 
problems, authors of \cite{KM92a} have introduced the non-local gauge
so that the gauge parameter in the gluon propagator becomes a momentum 
dependent function. On the other hand, we have proposed another way to 
recover the axial-vector WT identity \cite{NYNOTa}.  In this approach 
the axial-vector current is modified so as to become the correct Noether 
current of the effective model of QCD. 
The advantage of this approach is that it is 
applicable to all the effective models of QCD which respect the 
global chiral symmetry.
According to \cite{NYNOTa}, the meson decay constant can be 
expressed as follows\footnote{$f_\pi$ in this paper 
corresponds to $\tilde f_\pi$ in \cite{NYNOTa}.}.
\begin{eqnarray}
 f_B & = &\lim_{P\to P_B} \frac{1}{P^2} \int_q {\rm tr}\bigg[
 \overline{\chi}^R(q;P_B) \bigg\{ i\gamma_5\frac{\lambda^\alpha}{2} \bigg(
 \frac{f(-q_-^2)+f(-q_+^2)}{2}\Slash{P} + (f(-q_+^2)-f(-q_-^2))
 \Slash{q} \bigg) \nonumber \\
 & & \qquad\qquad\qquad\qquad\qquad\quad + E^\alpha(q;P) \bigg\} \bigg], 
 \label{AEQ:H100322:2}
\end{eqnarray}
\begin{eqnarray}
  E^\alpha_{mn}(q;P)  & := & \int_k \left[ \quad
  \left\{ {\cal K}^{n'n,mm'}
  \left(-k,q-\frac{P}{2};-q-\frac{P}{2},k+P \right) 
  \right. \right. \nonumber \\
 & & \quad\quad\quad
 - \left. {\cal K}^{n'n,mm'}\left(-k,q-\frac{P}{2};-q+\frac{P}{2},k \right)
  \right\}
 \left(i\gamma_5\frac{\lambda^\alpha}{2}S_F(k) \right)_{m'n'} \nonumber \\
 & & \quad\quad
 + \left\{ {\cal K}^{n'n,mm'}
   \left(-k+P,q-\frac{P}{2};-q-\frac{P}{2},k \right) \right. \nonumber \\
 & & \quad\quad\quad
 - \left. \left. {\cal K}^{n'n,mm'}
   \left(-k,q+\frac{P}{2};-q-\frac{P}{2},k \right) \right\}
   \left(S_F(k) i \gamma_5 \frac{\lambda^\alpha}{2}\right)_{m'n'} \right]
 \label{AEQ:H100322:3}
\end{eqnarray}
The on-shell value $f_B(M_E^2=-M_B^2)$
is obtained again by extrapolation from the space like region
$M_E^2>0$ to the on-shell point $M_E^2=-M_B^2$. For the neutral pion 
($B = \pi^0$), $\alpha = 3$ and so on in Eq.(\ref{AEQ:H100322:2}).
\par
     As shown in Ref.\cite{NYNOTa}, the WT identity for the axial-vector
vertex leads to the following relation in the 
improved ladder approximation model of QCD:
\begin{equation}
 M_B^2 f_B = -2m_R(\mu^2) {\cal E}_B(\mu^2)
 \label{AEQ:H100322:5}
\end{equation}
with
\begin{equation}
 {\cal E}_B(\mu^2) := 
 Z_m^{-1}(\mu^2) i \int_q  \frac{f(-q_{-}^2)+f(-q_{+}^2)}{2}{\rm tr} 
 \left[\overline{\chi}^R(q;P_B)\gamma_5\frac{\lambda^\alpha}{2}\right].
 \label{AEQ:H100808:1}
\end{equation}
This relation is satisfied for a finite quark mass.
In the case of chiral limit, we obtain
\begin{equation}
 f_B {\cal E}_B(\mu^2) = \xi_R(\mu^2)\quad \mbox{ where }\quad
 \xi_R := \langle \overline{\psi}\psi \rangle_R.
 \label{AEQ:H100322:6}
\end{equation} 
 One can treat Eq.$(\ref{AEQ:H100322:6})$ as an approximated relation
of the leading term of the expansion of $m_R$ for finite quark mass. 
 Substituting Eq.$(\ref{AEQ:H100322:6})$ to Eq.$(\ref{AEQ:H100322:5})$,
one obtains the Gell-Mann--Oakes--Renner (GMOR) mass formula
\begin{equation}
 M_B^2 f_B^2 \simeq -2m_R \langle \overline{\psi}\psi \rangle_R
|_{\mbox{\footnotesize chiral limit}}.
 \label{AEQ:H100520:4}
\end{equation}
\par
    For finite $m_R>0$, Eq.$(\ref{AEQ:H100322:5})$ is an
exact relation, while the violation of the GMOR formula
is incurred by the violation of Eq.$(\ref{AEQ:H100322:6})$.
\par
    We define ${\cal R}$ by 
\begin{equation}
 {\cal R}(M_E^2) := \frac{-M_E^2 f_B(M_E^2)}{-2m_R {\cal E}_B}. 
 \label{AEQ:H100408:1}
\end{equation}
A relation ${\cal R}(-M_B^2)=1$ must be satisfied due to 
Eq.$(\ref{AEQ:H100322:5})$.
We use this condition to make the extrapolation more reliable.
%
%
%
%
\section{Numerical results} \label{SEC:H100319:6}
%
%
\subsection{Parameters of the model}
    The parameters of the improved ladder model of QCD are the current 
quark mass $m_R$ for up and down quarks (The isospin symmetry is 
assumed throughout this paper.), the scale parameter of QCD 
$\Lambda_{\rm QCD}$, the infrared cut-off $t_{\rm IF}$ for the running 
coupling constant, the smoothness parameter $t_0$ and the ultraviolet
cut-off $\Lambda_{\rm UV}$.  We take $t_0 = -3$ throughout this paper, 
which is the same value used in Ref.\cite{KG1}.  In Ref.\cite{KG1}
they have shown that the numerical results are quite insensitive to 
the choice of the smoothness parameter in their studies of the BS amplitudes
in the chiral limit.  As for the ultraviolet cut-off $\Lambda_{\rm UV}$,
we shall show that the physical observables depends on it rather weakly
after our renormalization procedure described in section \ref{SEC:H100717:1},
if we use reasonably large value of $\Lambda_{\rm UV}$.  Of course, as we are 
treating not the full QCD but its approximation, we expect that weak 
dependences remain in our numerical studies.  Thus we only have three
physically relevant parameters, namely, the current
quark mass, $\Lambda_{\rm QCD}$ and the infrared cut-off.  
\par
    We choose $\Lambda_{\rm QCD}$ about $0.6$ [GeV].
It is rather large compared with the value obtained from the 
analyses in the deep inelastic scattering.
In the framework of this model, however,
one must employ the large value of $\Lambda_{\rm QCD}$ in order to
bring sufficiently strong dynamical chiral symmetry breaking.
It may be the indication of the limitation of the improved ladder approach. 
Other non-perturbative interactions between quarks may solve this discrepancy.
One candidate for such non-perturbative interactions is 
the instanton induced interaction proposed by 't~Hooft \cite{tHooft76}. 
There have been many studies of the roles of the instanton in low-energy 
QCD such as the instanton liquid model 
\cite{Shuryak82}, 
the generalized Nambu-Jona-Lasinio (NJL) model \cite{NJL} with 
't Hooft instanton induced interaction \cite{KH88},
the effects of the instanton in baryon sector 
\cite{SR89}.  The recent studies of the $\eta$-meson 
properties in the generalized NJL model with the 't~Hooft instanton 
induced interaction have shown \cite{TO95} that the contribution from the 
't~Hooft instanton induced interaction to the dynamical mass of the up and 
down quark mass is 44\% of that from the usual $U_L(3) \times U_R(3)$
invariant four-quark interaction.  The introduction of the 't~Hooft 
instanton induced interaction to the ILA model seems to be interesting and 
such attempt is now in progress \cite{NYNOT99}.
\par
\begin{figure}[tbp]
  \centerline{ \epsfxsize=10cm \epsfbox{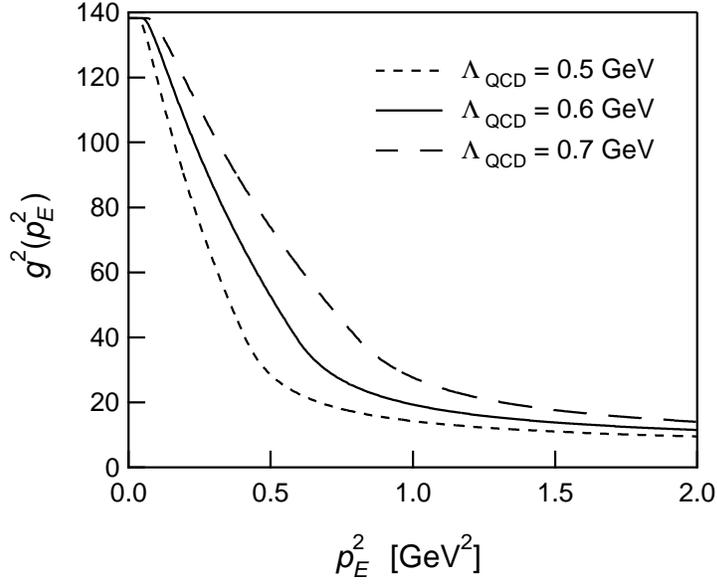} }
  \caption{$q^2_E$ dependence of $g^2(q_E^2)$ for various $\Lambda_{\rm QCD}$ 
  with $t_{\rm IF} = -0.5$.}
  \label{FIG:H100528:1}
\end{figure}
\begin{figure}[tbp]
  \centerline{ \epsfxsize=12cm \epsfbox{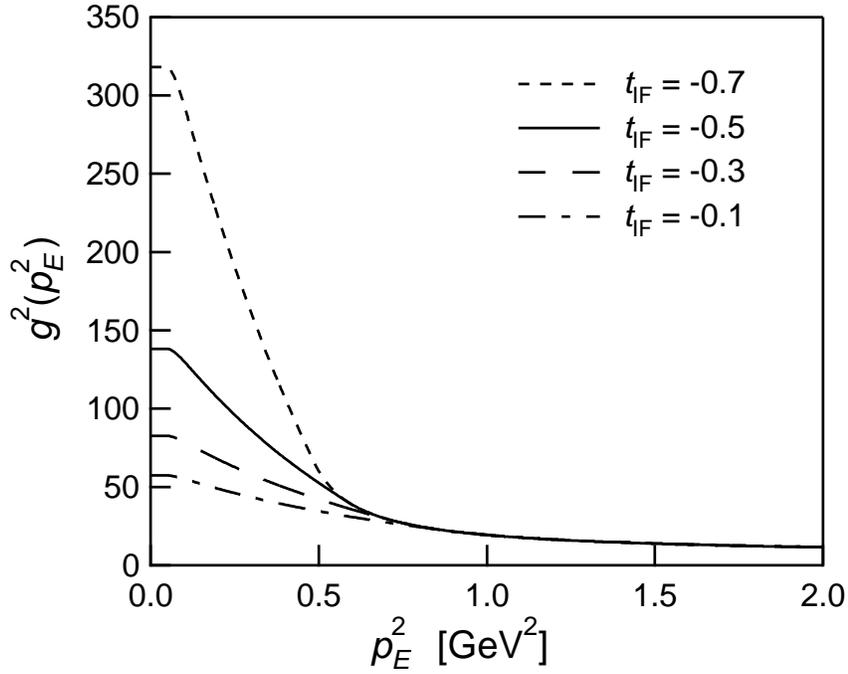} }
  \caption{$q^2_E$ dependence of $g^2(q_E^2)$ for various $t_{IF}$ with 
  $\Lambda_{\rm QCD}=0.6$GeV.}
  \label{FIG:H100528:2}
\end{figure}

    We employ $t_{\rm IF}$ about $-0.5$ due to Ref.\cite{KG1}. 
The running coupling constant for various $\Lambda_{\rm QCD}$ and 
$t_{\rm IF}$ is shown in Figs.\ref{FIG:H100528:1} and \ref{FIG:H100528:2}.
\subsection{SD equation}
\begin{figure}[tbp]
  \centerline{ \epsfxsize=12cm \epsfbox{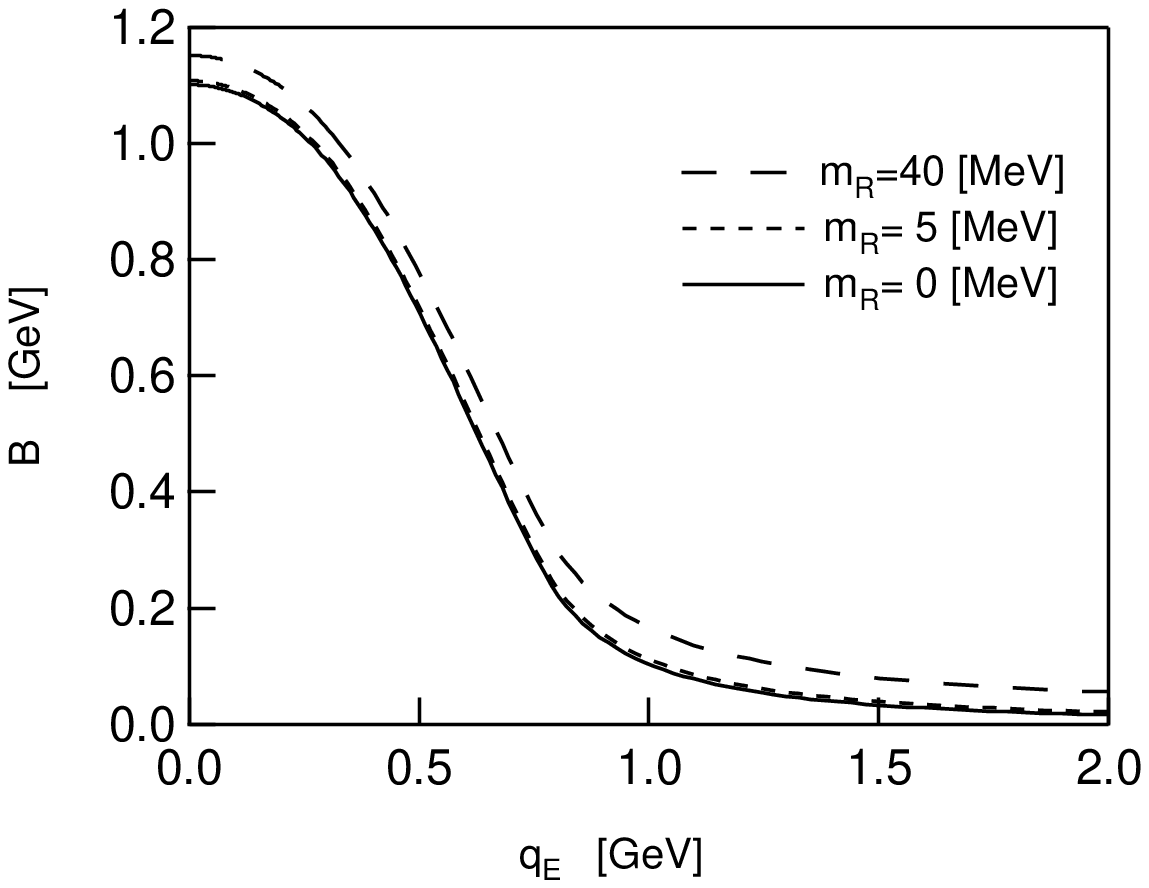} }
  \caption{Quark mass function $B(-q_E^2)$ as function of $q_E$.
  Model parameters are $\Lambda_{\rm UV}=100$ GeV, 
  $\Lambda_{\rm QCD}=0.6$ GeV, $t_{\rm IF}= -0.5$ and $\mu^2 = 4$ GeV$^2$.}
  \label{FIG:H100322:1}
\end{figure}

\begin{figure}[tbp]
  \centerline{ \epsfxsize=12cm \epsfbox{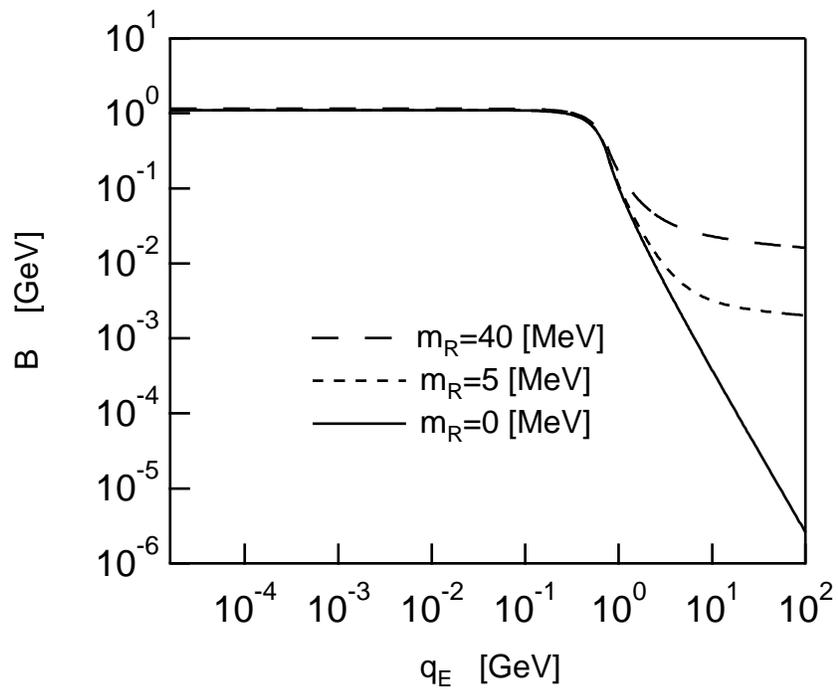} }
  \caption{Log-log plot of quark mass function $B(-q_E^2)$.
  Model parameters are $\Lambda_{\rm UV}=100$ GeV, 
  $\Lambda_{\rm QCD}=0.6$ GeV, $t_{\rm IF}= -0.5$ and $\mu^2 = 4$ GeV$^2$.}
  \label{FIG:H100922:1}
\end{figure}

    We discuss the solutions of the SD equation in this subsection.
$B(-q_E^2)$ as solutions of Eq.$(\ref{AEQ:H100319:24})$ for 
various values of $m_R$ are shown in Fig.\ref{FIG:H100322:1}.
In the chiral limit, i.e. $m_R=0$, we find a non-trivial 
solution $B(-q_E^2)$ which is non-zero for $q_E\le 1$GeV.
Note that Eq.$(\ref{AEQ:H100319:24})$ has also a
trivial solution $B(-q_E^2)=0$, if $m_R=0$. The existence of
the non-trivial solution indicates that chiral symmetry
is broken dynamically. $B(-q_E^2)$ decreases quickly to
zero for $q_E\ge 1$GeV. The asymptotic behavior is consistent with 
the OPE result as shown in Ref.\cite{H84}.
 The range of non-zero value of $B(-q_E^2)$ is determined by the 
$q_E^2$ dependence of the coupling constant, $g(q_E^2)$ defined in 
Eq.$(\ref{AEQ:H100319:8})$. The results shown in
Fig.\ref{FIG:H100322:1} correspond to the choice 
$\Lambda_{\rm QCD}=0.6$ GeV. When $\Lambda_{\rm QCD}$ decreases,
the range of $B(-q_E^2)$
decreases accordingly and therefore the chiral symmetry breaking is
weakened. But the $\Lambda_{\rm QCD}$ determines the scale of the system.
The order parameter $\langle \overline{\psi} \psi \rangle$ is almost
proportional to the $\Lambda^3_{\rm QCD}$ and the chiral symmetry
breaking always occurs for smaller $\Lambda_{\rm QCD}$. The parameter 
$t_{\rm IF}$ determines the strength of the coupling constant. 
Table \ref{TBL:H100527:1} shows $t_{\rm IF}$ dependence of 
the condensate in the chiral limit.
As can be seen from table \ref{TBL:H100527:1}, 
the chiral symmetry breaking does not occur for larger $t_{\rm IF}$.
\begin{table}[tbp]
\begin{center}
\begin{tabular}{|c|cccccc|} \hline
$t_{\rm IF}$ & $-0.5$ & $0.0$ & $0.5$ & $1.0$ & $1.5$ & $2.0$ \\ \hline
$ - \langle \overline{\psi}\psi \rangle_R^{1/3}$ [MeV] 
 & 259 & 240 & 187 & 116 & 43 & 2 \\ \hline 
\end{tabular}
\end{center}
\caption{$t_{\rm IF}$ dependence of quark condensate in the chiral limit.
Other model parameters are $\Lambda_{\rm UV}=100$ GeV, 
$\Lambda_{\rm QCD}=0.6$ GeV and $\mu^2 = 4$ GeV$^2$.}
\label{TBL:H100527:1}
\end{table}
The asymptotic behavior of $B(-q_E^2)$ with the finite current quark 
mass is rather different from that in the chiral limit.  
It can be seen clearly from the log-log plot of the 
quark mass function $B(-q_E^2)$ given in Fig. \ref{FIG:H100922:1}.
\par
\begin{figure}[tbp]
  \centerline{ \epsfxsize=12cm \epsfbox{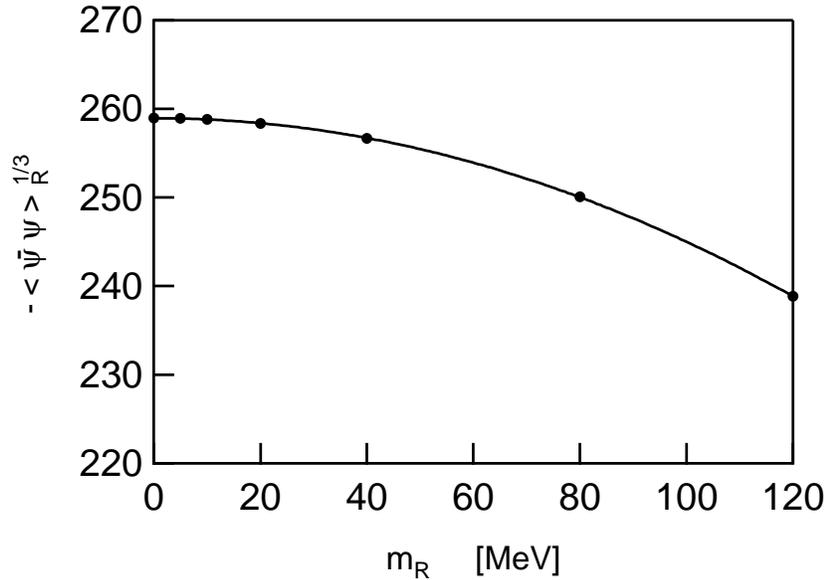} }
  \caption{$m_R$ dependence of 
  $-\langle \overline{\psi} \psi \rangle_R^{1/3}$. 
  Model parameters are $\Lambda_{\rm UV}=100$ GeV, 
  $\Lambda_{\rm QCD}=0.6$ GeV, $t_{\rm IF}= -0.5$ and $\mu^2 = 4$ GeV$^2$.}
  \label{FIG:H100923:2}
\end{figure}

    We next discuss the quark condensate. The quark condensates 
$-\langle \overline{\psi} \psi \rangle_R^{1/3}$ calculated for the 
various quark masses are shown in Fig. \ref{FIG:H100923:2}.  
Since in our definition of the quark condensate given in 
Eqs.(\ref{AEQ:H100722:4})-(\ref{AEQ:H100722:7}) the perturbative quark mass 
contribution is subtracted, $-\langle \overline{\psi} \psi \rangle_R^{1/3}$
decreases as $m_R$ increases.  Similar behavior is observed in the QCD 
sum rule approach. $\langle \overline{\psi} \psi \rangle_R$ 
for $m_R = 120$ MeV 
is about 78\% of $\langle \overline{\psi} \psi \rangle_R$ for $m_R = 5$ MeV. 
This is in reasonable agreement with the QCD sum rule result \cite{RRY85}: 
$\langle \bar ss \rangle / \langle \bar uu \rangle = 0.8 \pm 0.1$.
\subsection{BS equation}
\begin{figure}[tbp]
  \centerline{ \epsfxsize=12cm \epsfbox{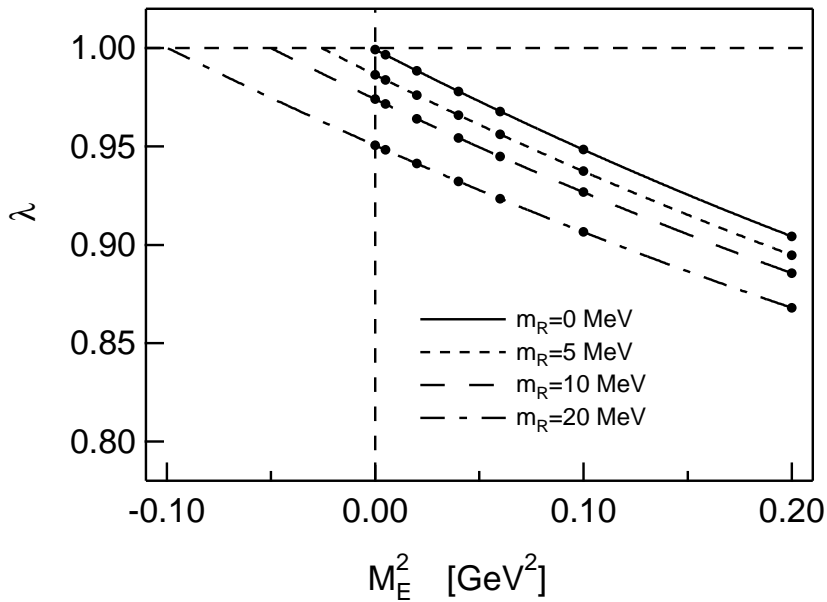} }
  \caption{Eigenvalues of BS equation $\lambda(M_E^2)$. 
  Model parameters are $\Lambda_{\rm UV}=100$ GeV, 
  $\Lambda_{\rm QCD}=0.6$ GeV, $t_{\rm IF}= -0.5$ and $\mu^2 = 4$ GeV$^2$.}
  \label{FIG:H100325:1}
\end{figure}

\begin{figure}[tbp]
  \centerline{ \epsfxsize=12cm \epsfbox{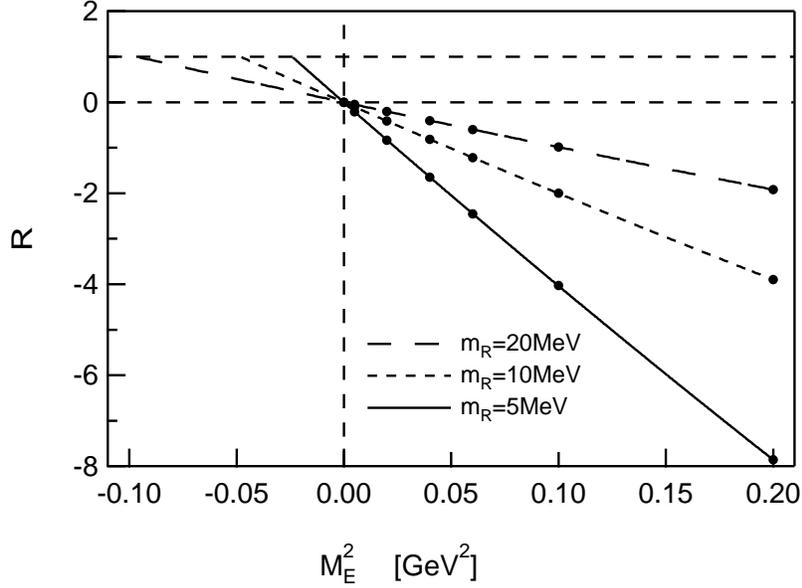} }
  \caption{The value ${\cal R}(M_E^2)$.
  Model parameters are $\Lambda_{\rm UV}=100$ GeV, 
  $\Lambda_{\rm QCD}=0.6$ GeV, $t_{\rm IF}= -0.5$ and $\mu^2 = 4$ GeV$^2$.}
  \label{FIG:H100408:1}
\end{figure}

\begin{figure}[tbp]
  \centerline{ \epsfxsize=12cm \epsfbox{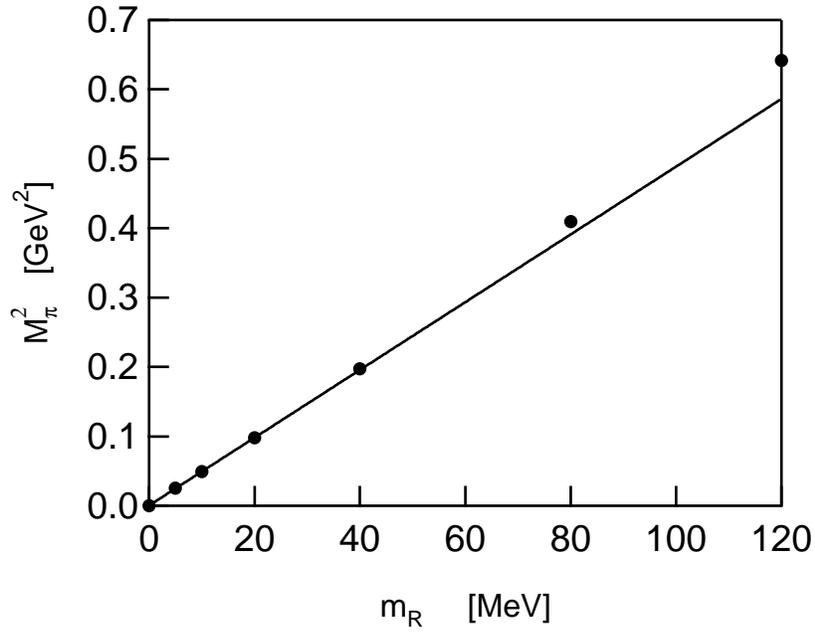} }
  \caption{ $M_\pi^2$ as a function of $m_R$. Dots represent our
  calculated results and line represents the linear fit of the first 4 points.
  Model parameters are $\Lambda_{\rm UV}=100$ GeV, 
  $\Lambda_{\rm QCD}=0.6$ GeV, $t_{\rm IF}= -0.5$ and $\mu^2 = 4$ GeV$^2$.}
  \label{FIG:H100507:1}
\end{figure}

    Let us now turn to the discussion of the solutions of the BS equation.
 The eigenvalues $\lambda(M_E^2)$ of the BS equation are shown 
in Fig.\ref{FIG:H100325:1}.
One sees that the massless solution $\lambda(M_E^2=0)=1$ 
appears in the chiral limit.
This is a result of the Nambu--Goldstone theorem.
For a non-zero quark mass, we need to extrapolate $\lambda(M_E^2)$
to the time-like $M^2_E<0$.  We have fitted $\lambda(M_E^2)$ by a
quadratic function using the method of least-squares
and extrapolated $\lambda(M_E^2)$ to the time-like region to find the 
point at which $\lambda(M_E^2)$ becomes unity.
Fig.\ref{FIG:H100325:1}
implies that the extrapolation length is longer for larger quark mass.
In order to reduce the ambiguity in the extrapolation procedure, we also
evaluate the ratio $\cal R$ defined by Eq.$(\ref{AEQ:H100408:1})$ as a
function of $M_E^2$. Because of the exact relation $(\ref{AEQ:H100322:5})$,
$\cal R$ must hit ${\cal R}(-M_\pi^2)=1$ at the on-mass-shell of the pion.
The value ${\cal R}$ is shown in Fig.\ref{FIG:H100408:1}.
We fit the graph with the linear function using the methods
of least-squares. 
We show our calculated results of the pion mass $M_\pi$ 
determined by the above mentioned two conditions, i.e., 
$\lambda = 1$ and ${\cal R} = 1$, 
for various values of the quark mass $m_R$ in table \ref{TBL:H100507:1}.
The ambiguity by the extrapolation procedure is reasonably small up to 
the strange quark mass region.
\begin{table}[tbp]
\begin{center}
\begin{tabular}{|c|ccccccc|} \hline
 $m_R$ & 0 & 5 & 10 & 20 & 40 & 80 & 120 \\ \hline
 $M_\pi$ $(\lambda=1)$ & 0.0 & 159.1 & 222.0 & 312.9 & 444.4 & 639.8 & 800.9 \\ \hline
 $M_\pi$ $({\cal R}=1)$ & 0.0 & 154.5 & 218.5 & 309.0 & 436.9 & 616.1 & 749.7\\ \hline
\end{tabular}
\end{center}
\caption{ Pion masses determined by the two conditions, $\lambda = 1$ and
${\cal R}=1$ for various values of $m_R$. 
All the entries are in units of MeV.
Model parameters are 
$\Lambda_{\rm UV}=100$ GeV, $\Lambda_{\rm QCD}=0.6$ GeV, 
$t_{\rm IF}= -0.5$ and $\mu^2 = 4$ GeV$^2$.}
\label{TBL:H100507:1}
\end{table}
We plot our results of $M_\pi^2$ 
obtained by the condition $\lambda = 1$ as a function of $m_R$ 
in Fig. \ref{FIG:H100507:1}.
The $M_\pi^2$ seems to be almost a linear function of $m_R$ up to 
$m_R \sim 40$ MeV.
This is suggested by the GMOR formula
\begin{equation}
 M_\pi^2 \simeq \left(\frac{ -2\langle
 \overline{\psi}\psi 
 \rangle_R }{f_\pi^2}\right)_{\mbox{\footnotesize chiral limit}} m_R \, .
 \label{AEQ:H100507:2}
\end{equation}
The deviation from the linear dependence at $m_R = 120$ MeV is about 9\%.
\par
\begin{figure}[tbp]
  \centerline{ \epsfxsize=12cm \epsfbox{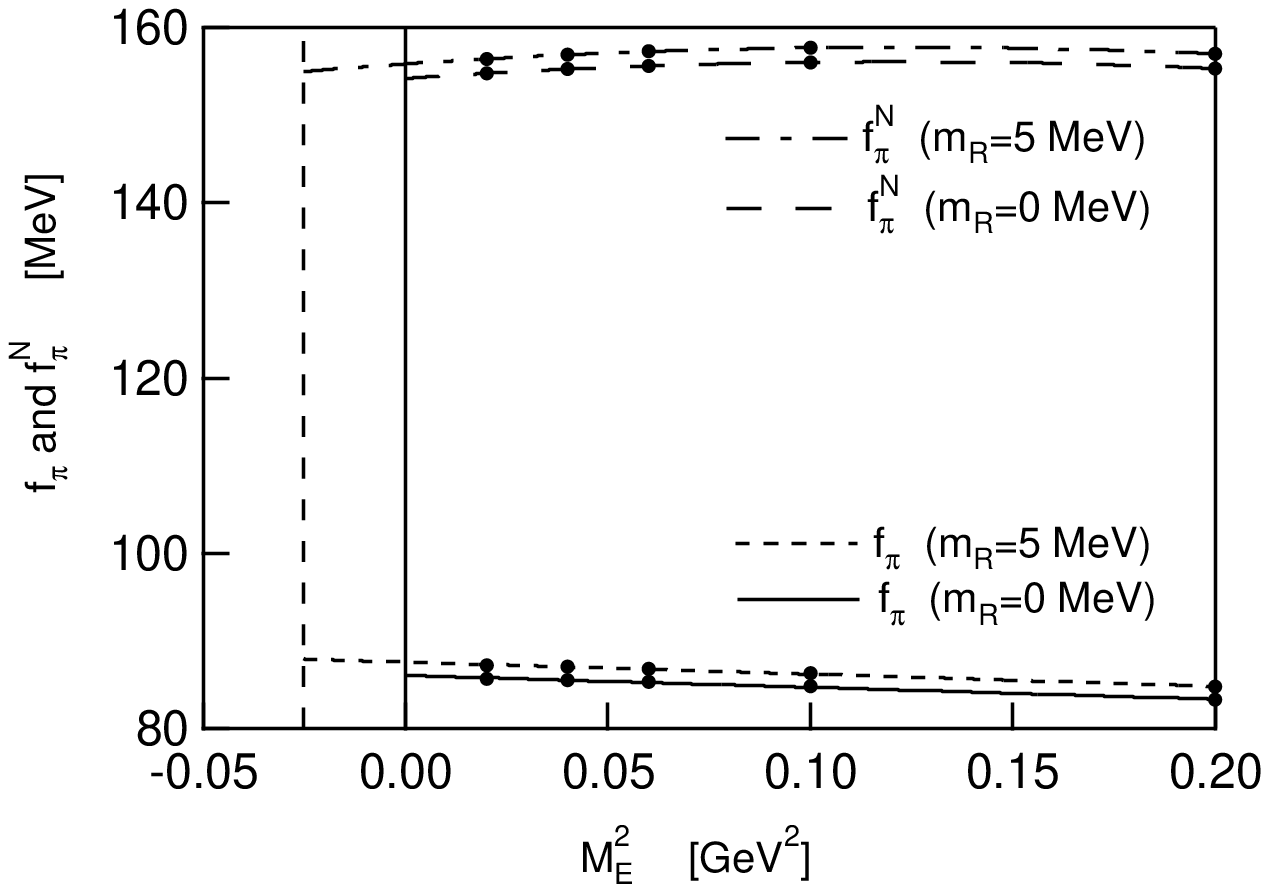} }
  \caption{$M_E^2$ dependence of $f_\pi$ and $f_\pi^N$.  Vertical dashed 
  line represents the point $M_E^2 = -M_\pi^2$ for $m_R =5$ MeV. 
  Model parameters are $\Lambda_{\rm UV}=100$ GeV, 
  $\Lambda_{\rm QCD}=0.6$ GeV, $t_{\rm IF}= -0.5$ and $\mu^2 = 4$ GeV$^2$.}
  \label{FIG:H100407:1}
\end{figure}

    We next discuss the pion decay constant.  As mentioned in 
Sec. \ref{SEC:H100319:5}, we can calculate $f_\pi(M_E^2)$ only 
for the time-like $M_E^2$ and therefore the on-shell value of the 
decay constant $f_\pi(M_E^2 = - M_\pi^2)$ can be obtained again by the 
extrapolation.  The $M_E^2$ dependence of the pion decay constants  
in the chiral limit and in the case of $m_R = 5$ MeV are shown 
in Fig. \ref{FIG:H100407:1}.
\par
  To estimate the effect of $E^\alpha(q;P)$ in Eq.$(\ref{AEQ:H100322:2})$,
we plot the naive value 
$f^N_\pi$ which is defined by neglecting $E^\alpha(q;P)$ term from 
Eq.$(\ref{AEQ:H100322:2})$.
It seems to be a good approximation that $f_\pi(M_E^2)$ is a linear
function of $M_E^2$.  Therefore we fit the curve by the linear
function using the method of least-squares and make an extrapolation 
to the time-like $M_E^2$ for finite $m$ \footnote{
It should be noted that the decay constant at 
small (positive) $M_E^2$ suffers from numerical uncertainty and 
thus it deviates from the straight line.  We do not use these points 
in our extrapolation procedure.}. 
On the other hand, we fit 
$f_\pi^N$ by the quadratic function using the least-square method to
extrapolate to the on-shell point $M_E^2 = M_\pi^2$. 
Our results for $m_R = 5$ MeV are 
$f_\pi = 88$ MeV and $f_\pi^N = 155$ MeV, so the contribution of 
$E^\alpha(q;P)$ term is remarkable.
Similar result has been found in the ILA model which respects the 
axial-vector WT identity by using the gluon momentum square as the 
argument of the running coupling constant and the non-local gauge 
\cite{KM92a}.
The condition ${\cal R}(-M_\pi^2) = 1$ is the direct consequence of 
the axial-vector WT identity and therefore it has been proved 
numerically that our definition of the decay constant is consistent 
with the axial-vector WT identity from the fact that the pion mass 
determined by the condition $\lambda = 1$ is almost same as that 
determined by the condition ${\cal R} = 1$.
\par
\begin{figure}[tbp]
  \centerline{ \epsfxsize=12cm \epsfbox{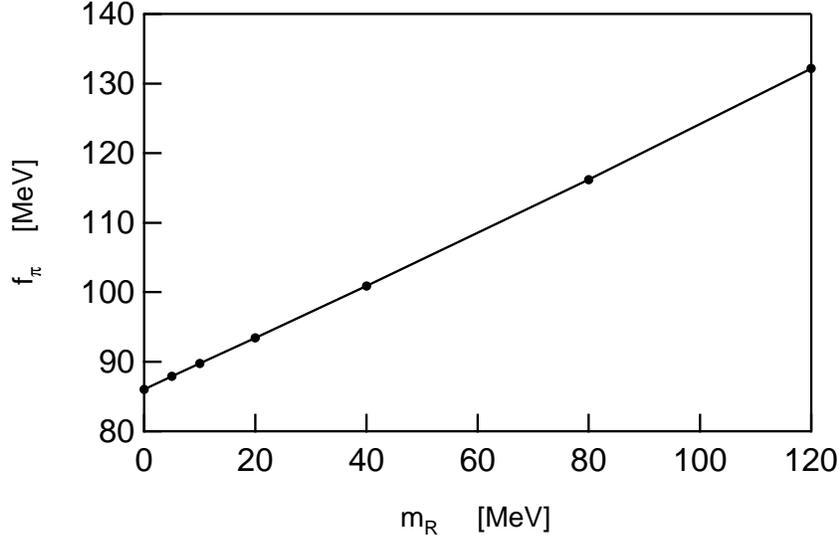} }
  \caption{$m_R$ dependence of $f_\pi$.   
  Model parameters are $\Lambda_{\rm UV}=100$ GeV, 
  $\Lambda_{\rm QCD}=0.6$ GeV, $t_{\rm IF}= -0.5$ and $\mu^2 = 4$ GeV$^2$.}
  \label{FIG:H100923:1}
\end{figure}

    We plot the quark mass dependence of the decay constant in 
Fig. \ref{FIG:H100923:1}.
$f_\pi$ almost linearly depends on $m_R$ in our case.  $f_\pi$ at $m_R = 40$ 
MeV is about 15\% bigger than $f_\pi$ at $m_R = 5$ MeV and 
$f_\pi$ at $m_R = 80$ MeV is about 32\% bigger than $f_\pi$ at $m_R = 5$ MeV.
Since the observed $f_K/f_\pi = 1.23$, $m_R$ dependence of the decay constant 
seems to be reasonable though we have not solved the kaon BS equation. 
This $m_R$ dependence of $f_\pi$ is similar to that obtained in the 
chiral perturbation theory (ChPT) \cite{GL84} though the chiral log term
has not been seen in our numerical result.
It is understandable because the Goldstone boson loop contribution is not 
taken into account explicitly in our approach.
On the other hand, the quark-antiquark structure is included explicitly and 
the finite quark mass effects are fully taken into account without
performing the perturbative expansion with respect to the quark mass.
\begin{table}[tbp]
\begin{center}
\begin{tabular}{|c|c|c|c|} \hline
  $\Lambda_{\rm QCD}$ [GeV] & $M_\pi$ [MeV] & $f_\pi$ [MeV] &
  $-\langle \overline{\psi} \psi \rangle^{1/3}_R$ [MeV]\\ \hline
  0.5 & 152 & 74 & 223 \\
  0.6 & 159 & 88 & 259 \\
  0.7 & 166 & 102 & 293 \\ \hline
\end{tabular}
\end{center}
\caption{$\Lambda_{\rm QCD}$ dependences of $M_\pi$, $f_\pi$ and 
         $-\langle \overline{\psi} \psi \rangle^{1/3}_R$. 
         Other model parameters are $\Lambda_{\rm UV}=100$ GeV, 
         $t_{\rm IF}= -0.5$, $\mu^2 = 4$ GeV$^2$ and $m_R = 5$ MeV.}
\label{TBL:H100416:1}
\end{table}
\begin{table}[tbp]
\begin{center}
\begin{tabular}{|c|c|c|c|} \hline
  $t_{\rm IF}$ & $M_\pi$ [MeV] & $f_\pi$ [MeV] & 
  $-\langle \overline{\psi} \psi \rangle^{1/3}_R$ [MeV] \\ \hline
$-0.3$ & 149 & 91 & 256 \\
$-0.5$ & 159 & 88 & 259 \\
$-0.7$ & (181) & (74) & (253) \\ \hline
\end{tabular}
\end{center}
\caption{$t_{\rm IF}$ dependences of $M_\pi$, $f_\pi$ and 
         $-\langle \overline{\psi} \psi \rangle^{1/3}_R$. 
         Other model parameters are $\Lambda_{\rm UV}=100$ GeV, 
         $\Lambda_{\rm QCD} = 0.6$ GeV, $\mu^2 = 4$ GeV$^2$ and $m_R = 5$ MeV.}
\label{TBL:H100428:1}
\end{table}
\par
    Let us now discuss the $\Lambda_{\rm QCD}$ and the $t_{\rm IF}$ 
dependences.
Table \ref{TBL:H100416:1} shows the $\Lambda_{\rm QCD}$ dependences of 
the pion mass, the pion decay constant and the quark condensate.
As shown in Table \ref{TBL:H100416:1}, the all the quantities with the mass 
dimension one are roughly proportional to the $\Lambda_{\rm QCD}$.
It is understandable since the only scale of the theory is the 
$\Lambda_{\rm QCD}$ if one can neglect the current quark mass. 
Table \ref{TBL:H100428:1} shows the $t_{\rm IF}$ dependence.
For $t_{\rm IF}$ below $-0.7$ the coupling constant becomes
very steep and our numerical procedure is not sufficiently accurate.
Although we have not performed the fine tuning of the model parameters, 
it is clear from  
Tables \ref{TBL:H100527:1}, \ref{TBL:H100416:1} and \ref{TBL:H100428:1} 
that one can fit the model parameters so as to reproduce the observed 
values of $M_\pi$ and $f_\pi$ and the empirically determined value of 
the quark condensate. 
\par
   We study the $\Lambda_{\rm UV}$ dependence by changing the value of 
$\Lambda_{\rm UV}$ from 10 GeV to 1000 GeV.  It causes less than 1\% 
changes of the $M_\pi$, $f_\pi$ and $\langle \overline{\psi} \psi \rangle_R$.
This stability indicates that our non-perturbative renormalization 
procedure works well. 
\subsection{Approximation}
\begin{figure}[tbp]
  \centerline{ \epsfxsize=12cm \epsfbox{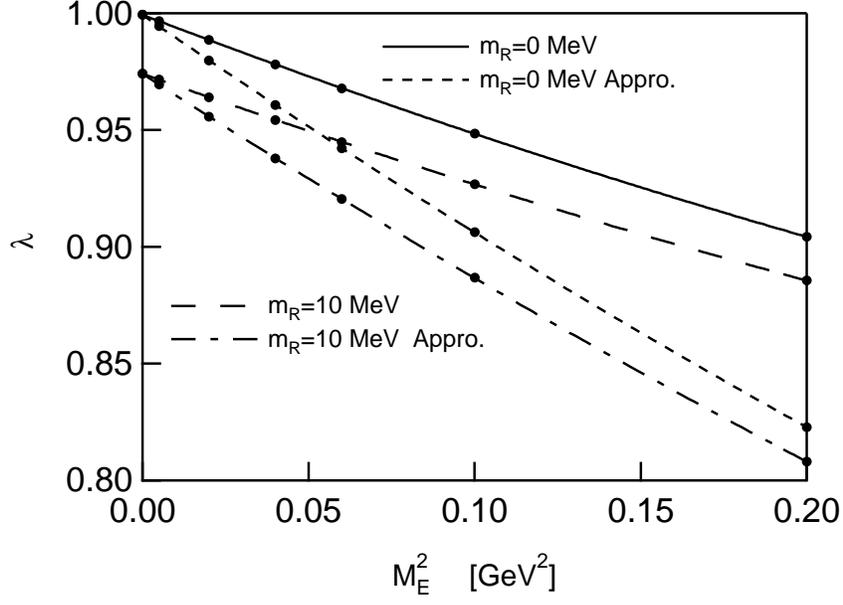} }
  \caption{Eigenvalues of the BS equations with and without 
  the approximation described in the text. 
  Model parameters are $\Lambda_{\rm UV}=100$ GeV, 
  $\Lambda_{\rm QCD}=0.6$ GeV, $t_{\rm IF}= -0.5$ and $\mu^2 = 4$ GeV$^2$.}
  \label{FIG:H100416:1}
\end{figure}

    Finally we discuss the approximation often used in solving the 
BS equation.
 The approximation in which one neglects the
$\phi_P,\phi_Q,\phi_T$ terms in RHS in the BS equation 
$(\ref{AEQ:H100320:6})$
is often used in literatures \cite{FR96}.
The resulting eigenvalues are shown 
in Fig.\ref{FIG:H100416:1}.
While this approximation gives the massless NG boson in the chiral limit, 
it underestimates the pion mass 
for finite quark mass.
The decay constants obtained from the approximated BS amplitude 
are shown in Table \ref{TBL:H100423:1}.
\begin{table}[tbp]
\begin{center}
\begin{tabular}{|c|c|c|c|c|c|c|} \hline
$m_R$ & $M_\pi$ & $M_\pi^{\rm Appro.}$ & $f_\pi$ & $f_\pi^{\rm Appro.}$ 
& $f_\pi^N$ & $f_\pi^{N {\rm Appro.}}$ \\ \hline
0 & 0 & 0 & 86 & 115 & 154 & 116 \\ 
5 & 159 & 119 & 88 & 117 & 156 & 117 \\ 
10 & 222 & 166 & 90 & 118 & 158 & 119 \\ \hline
\end{tabular}
\end{center}
\caption{Result with and without Approximation. 
  All the entries are in units of MeV.
  Model parameters are $\Lambda_{\rm UV}=100$ GeV, 
  $\Lambda_{\rm QCD}=0.6$ GeV, $t_{\rm IF}= -0.5$ and $\mu^2 = 4$ GeV$^2$.}
\label{TBL:H100423:1}
\end{table}
The approximation overestimates the pion decay constant about 30\%
while the axial-vector WT identity 
$(M_\pi^{\rm Appro.})^2 f_\pi^{\rm Appro.} 
= - 2 m_R {\cal E}_\pi^{\rm Appro.}$
is preserved.  It is seen that the effect of $E^\alpha(q;P)$
is very small in this case, and 
$f_\pi^{\rm Appro.} \simeq f_\pi^{N {\rm Appro.}}$.
Therefore, the violation of the axial-vector WT
identity or that of the exact relation for the PCAC current
incured by neglecting $E^\alpha(q;P)$ effect is very small.
\par
  The above $f_\pi^{\rm Appro.}$ is calculated in a similar way
as in the approximation discussed in Ref.\cite{NYNOTa,NO98}.
The following is shown in theorem 2 of \cite{NO98}.
If the interaction is local chiral invariant, the approximation of
taking the wave function renormalization of the quark propagator to be one 
and at the same time neglecting the $\phi_P$, $\phi_Q$ and $\phi_T$ terms 
in the RHS of the BS equation preserves the low-energy relations.
This theorem cannot be applied to the present case since the interaction 
term of the ILA model 
breaks the local chiral symmetry.  However one can prove that 
the low-energy relation holds if one neglects the $\phi_P$ and $\phi_Q$ 
terms in $E^\alpha(q;P)$ as well as the $\phi_P$, $\phi_Q$ and $\phi_T$ terms 
in the RHS of the  BS equation by following the same argument of the proof 
of theorem 2 in \cite{NO98}.
%
%
%
%
\section{Conclusion} \label{SEC:H100319:7}
 We have solved the Schwinger--Dyson (SD) equation for the quark propagator 
and the Bethe--Salpeter (BS) equation for the pion in the improved ladder 
approximation of QCD. 
We have carefully treated the consistency of the equations in order 
to preserve the low-energy relations associated with chiral symmetry
by using the Cornwall--Jackiw--Tomboulis effective action approach. 
We have introduced the finite quark mass term in order to study effects of
explicit chiral symmetry breaking on the low-energy relations. 
Because of the difference in the asymptotic behavior of
the quark mass function for finite quark mass, the non-perturbative 
mass-independent renormalization has been introduced and the quark 
condensate for finite quark mass is calculated.  
In solving the SD and BS equations, we have not 
taken any further approximation such as expansion of BS amplitudes 
in the Gegenbauer polynomials. 
\par
   We have obtained reasonable values of $M_\pi$, $f_\pi$ and 
$\langle \overline{\psi} \psi \rangle_R$ with a rather large value of 
$\Lambda_{\rm QCD}$.  It may indicate the limitation of the improved ladder
approach.
The pion mass $M_\pi$ grows as quark mass $m_R$ increases.
Up to the strange quark mass region $M_\pi^2$ seems to be proportional to 
quark mass $m_R$ almost as predicted by the GMOR relation
\begin{equation}
 M_\pi^2 = \left( \frac{- 2 \langle \overline{\psi}\psi \rangle_R}
                    {f_\pi^2} \right)_{\mbox{\footnotesize chiral limit}} m_R.
 \label{AEQ:H100423:1}
\end{equation}
We have found that the $f_\pi$ also grows as $m_R$ increases almost linearly. 
The $m_R$ dependences of $M_\pi^2$ and $f_\pi$ are similar to those obtained 
in the chiral perturbation theory.
It suggests that the chiral perturbation is applicable up to the strange quark
mass region.  
\par
   We have studied the effect of $E^\alpha(q;P)$ term in the
true decay constant.  We have found that it is significantly large for various 
input parameters. Therefore in the framework of the improved ladder 
approximation, $E^\alpha(q;P)$ plays an essential role to keep the 
chiral property.
\par
   We have further shown the result of the approximation neglecting 
$\phi_P(q;P),\phi_Q(q;P)$ and $\phi_T(q;P)$ term in RHS of the BS equations.
This approximation is very useful and makes the calculation easy greatly.
But the result gives a smaller pion mass.
This suggests that the simple picture of the $\phi_S(q;P)$ dominance 
in the BS equation is not so good, at least in the present model.
\par
   So far, we have studied the symmetric $q$-$\bar q$ systems,
$u\bar u$, $d\bar d$, etc. It is interesting to extend the present 
formulation to asymmetric systems like the kaon. It is also interesting
to introduce the $U_A(1)$ breaking interaction to this framework and
to study the $\eta$-$\eta'$ systems. Such attempts are in progress. 
%
%
%
%
\section*{Acknowledgment}
    The authors would like to thank Kensuke Kusaka for useful discussions.
This work is supported in part by the Grant-in-Aid for Scientific Research
(A)(1)08304024 and (C)(2)08640356 of the Ministry of Education, Science,
Sports and Culture of Japan.  
%
%
%
%
\appendix
\section*{Appendix}
 Here we write down the BS equation explicitly.
 In this section the total momentum is denoted by $P$ instead by $P_B$ 
for simplicity.  
 First we define the regularized amputated BS amplitude 
$\hat{\chi}^R(q;P)$ by
\begin{equation}
 \hat{\chi}^R(q;P) := S_F^{R-1}(q+\frac{P}{2}) \chi^R(q;P) 
 S_F^{R-1}(q-\frac{P}{2}) , \label{AEQ:H100528:2}
\end{equation}
which can be expressed in terms of 
\begin{eqnarray}
 \lefteqn{ 
 \hat{\chi}^R_{nm}(q;P) = \delta_{ji} \frac{(\lambda^a)_{gf}}{2}
 \bigg[  \bigg(\hat{\phi}_S(q;P) + \hat{\phi}_P(q;P) 
 \Slash{q} + \hat{\phi}_Q(q;P)\Slash{P} 
  } \nonumber \\
  & & \quad {}  +\frac{1}{2}\hat{\phi}_T(q;P)(\Slash{P}
  \Slash{q}-\Slash{q}\Slash{P})\bigg)\gamma_5\bigg]_{ba} .
 \label{AEQ:H100528:3}
\end{eqnarray}
The BS equation $(\ref{AEQ:H100320:6})$ reads
\begin{equation}
 \hat{\phi}_{\cal A}(q;P) = \int_k K_{{\cal AB}}(q,k;P)
  \phi_{\cal B}(k;P).
 \label{AEQ:H100528:4}
\end{equation}
The components of the kernel is given explicitly by
\begin{eqnarray}
 K_{SS}(q,k;P) & = & iC_F \bar{g}^2(q,k)\frac{-3}{(q-k)^2}
 \label{AEQ:H100720:1} \\
 K_{PP}(q,k;P) & = & \frac{iC_F  \bar{g}^2(q,k)}{P^2q^2-(Pq)^2} 
  \bigg\{ \frac{P^2(qk)-(Pq)(Pk)}{(q-k)^2} \nonumber \\
 & & {} + \frac{2(qk-k^2)(P^2q^2-(Pq)^2+(Pq)(Pk) 
 - P^2(qk))}{(q-k)^4} \bigg\}
 \label{AEQ:H100720:2} \\
 K_{PQ}(q,k;P) & = & \frac{iC_F\bar{g}^2(q,k)}{P^2q^2-(Pq)^2} 
 \cdot \frac{2(Pq-Pk)(P^2q^2-(Pq)^2+(Pq)(Pk)-P^2(qk))}{(q-k)^4} 
 \label{AEQ:H100720:3} \\
 K_{QP}(q,k;P) & = & \frac{iC_F \bar{g}^2(q,k)}{P^2q^2-(Pq)^2}
 \bigg\{ \frac{(Pk)q^2-(Pq)(qk)}{(q-k)^2} \nonumber \\
 & & {} + \frac{2(qk-k^2)((Pq)(qk)-(Pk)q^2)}{(q-k)^4} \bigg\}
 \label{AEQ:H100720:4} \\
 K_{QQ}(q,k;P) & = & \frac{iC_F\bar{g}^2(q,k)}{P^2q^2-(Pq)^2}
 \bigg\{ \frac{P^2q^2-(Pq)^2}{(q-k)^2} \nonumber \\
 & & {} + \frac{2(Pq-Pk)((Pq)(Pk)-(qk)q^2)}{(q-k)^4} \bigg\}
 \label{AEQ:H100720:5} \\
 K_{TT}(q,k;P) & = & \frac{iC_F \bar{g}^2(q,k)}{P^2q^2-(Pq)^2} \cdot
 \frac{1}{(q-k)^4} \bigg\{ (k^2-q^2)((Pq)(Pk)-P^2(qk)) \nonumber \\
 & & {} + 2(Pq-Pk)((Pk)q^2-(Pq)(qk))-2(qk-k^2)(P^2q^2-(Pq)^2)\bigg\}
 \label{AEQ:H100720:6}
\end{eqnarray}
and other components are zero.
The relations between $\phi_{\cal A}(q;P)$ and $\hat{\phi}_{\cal A}(q;P)$ 
are given by
\begin{eqnarray}
 \phi_S(q;P) & = & \frac{1}{\Delta} \Bigg[ \left\{ q^2-\frac{P^2}{4}-B(q_+^2)B(q_-^2) \right\} \hat{\phi}_S(q;P) \nonumber \\
 & & {} + \left\{ q^2(B(q_+^2)-B(q_-^2)) - \frac{Pq}{2} (B(q_-^2)+B(q_+^2)) \right\} \hat{\phi}_P(q;P) \nonumber \\
 & & {} + \left\{ (Pq)(B(q_+^2)-B(q_-^2)) - \frac{P^2}{2}(B(q_-^2)+B(q_+^2)) \right\} \hat{\phi}_Q(q;P) \nonumber \\
 & & {} + (P^2q^2-(Pq)^2) \hat{\phi}_T(q;P) \Bigg] 
 \label{AEQ:H100720:7} \\
 \phi_P(q;P) & = & \frac{1}{\Delta} \Bigg[ 
 (B(q_+^2)-B(q_-^2)) \hat{\phi}_S(q;P) \nonumber \\
 & & {} + \left\{ q^2+\frac{P^2}{4} - B(q_+^2)B(q_-^2) \right\} \hat{\phi}_P(q;P) \nonumber \\
 & & {} + 2(Pq) \hat{\phi}_Q(q;P) \nonumber \\
 & & {} - \left\{ (Pq)(B(q_-^2)+B(q_+^2)) 
 + \frac{P^2}{2}(B(q_-^2)-B(q_+^2)) \right\} \hat{\phi}_T(q;P) 
 \Bigg] \label{AEQ:H100720:8} \\
 \phi_Q(q;P) & = & \frac{1}{\Delta} \Bigg[
 -\frac{1}{2}(B(q_+^2)+B(q_-^2)) \hat{\phi}_S(q;P) \nonumber \\
 & & {} - \frac{Pq}{2} \hat{\phi}_P(q;P) \nonumber \\
 & & {} -\left\{ q^2+\frac{P^2}{4}+B(q_+^2)B(q_-^2) \right\} 
 \hat{\phi}_Q(q;P) \nonumber \\
 & & {} +\left\{ q^2(B(q_-^2)+B(q_+^2)) 
 + \frac{Pq}{2}(B(q_-^2)-B(q_+^2)) \right\} \hat{\phi}_T(q;P) 
 \Bigg] \label{AEQ:H100720:9} \\
 \phi_T(q;P) & = & \frac{1}{\Delta} \Bigg[
 \hat{\phi}_S(q;P) -\frac{1}{2}(B(q_-^2)-B(q_+^2)) \hat{\phi}_P(q;P) \nonumber \\
 & & {} + (B(q_+^2)+B(q_-^2)) \hat{\phi}_q(q;P) \nonumber \\
 & & {} + \left\{ -q^2+\frac{P^2}{4}-B(q_-^2)B(q_+^2)\right\} \hat{\phi}_T(q;P) \Bigg] \label{AEQ:H100720:10}
\end{eqnarray}
where
\begin{equation}
 \Delta  := (q_+^2-B^2(q_+^2)) (q_-^2-B^2(q_-^2)).
 \label{AEQ:H100720:11}
\end{equation}
\end{document}